# A NEW INNER DRIFT CHAMBER FOR BESIII MDC

M.Y. DONG*, Z.H. QIN, X.Y. MA, J. ZHANG, J. DONG, W. XIE, Q.L. XIU, X.D. JU, R.G. LIU and Q.OUYANG

*Experiment Physics Center, Institute of High Energy Physics, Chinese Academy of Sciences*
*Beijing, 100049, China*
*\*dongmy@ihep.ac.cn*

Due to the beam related background, the inner chamber of BESIII MDC has aging effect after 5 years running. The gains of the inner chamber cells decrease obviously, and the max gain decrease is about 26% for the first layer cells. A new inner drift chamber with eight stereo sense wire layers as a backup for MDC is under construction, which is almost the same as the current one but using stepped endplates to shorten the wire length beyond the effective solid angle. This new structure will be of benefit to reducing the counting rate of single cell. The manufacture of each component is going smoothly, and the new inner drift chamber will be finished by the end of April 2014.

## 1. Introduction

The Beijing spectrometer III (BESIII) is a high precision detector for the Beijing electron positron collider II (BEPCII), a high luminosity, multi-bunch $e^+ e^-$ collider running at the tau-charm energy region[1]. As its center tracking detector, BESIII main drift chamber (MDC) is a low-mass cylindrical chamber with small-cell geometry, using helium-based gas and operating in a 1T magnetic field. MDC consists of an inner chamber and an outer chamber, which are joined together at the endplates, sharing a common gas volume[2]. The inner chamber was designed to be replaced in case of radiation damage.

After it has been running five years, the inner chamber of BESIII MDC is suffering from the aging problems because of the huge beam related background. The gains of the inner chamber have an obvious decrease, and the gain of the inner most layer decrease a maximum of about 26%. In addition, in order to reduce the dark currents of the sense wires to protect the detector, the operation high voltages of the first four layers have to be set to 96%, 97%, 98% and 99% of the normal value respectively, which results in the decrease of the effective gain and leads to the performance of the inner chamber become bad. Besides the study on upgrading the inner chamber with some new technologies, such as cylindrical gas electron multiplier (GEM), CMOS pixel sensor, building a backup of inner chamber is to be accomplished firstly. In this paper, we report the design and construction of the new inner drift chamber.

## 2. The new inner drift chamber

The new inner drift chamber is composed of an inner carbon fiber cylinder and two multi-stepped endplates, including eight stereo sense wire layers. The length of the new inner chamber is 1092 mm, and the radial extent is from 59 mm to 183.5 mm, which is almost the same as the current one. In order to reduce the background counting rate, new multi-stepped endplates are adopted to shorten the wire length beyond the effective detection solid angle, as shown in Fig. 1. This new structure will be of benefit to reducing the dark current of single cell. The wire length of the new inner chamber compared with the current one is shown in table1, and the expected reduction of the background counting rate is in the range from 31% to 4% for each layer.

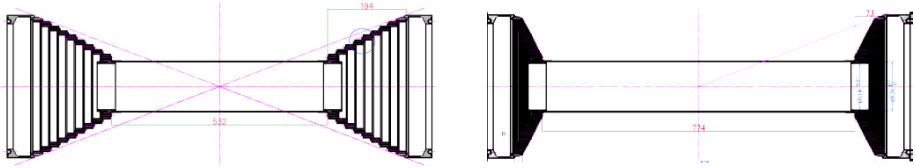

Fig. 1. The structure of the new inner drift chamber (left) and the current one (right).

Table 1. The wire length of the new inner chamber compared with the current one.

| No. of wire layer | Current wire length (cm) | New wire length (cm) | Background rate reduction expected |
|---|---|---|---|
| 1 | 78.0 | 53.8 | 31% |
| 2 | 79.2 | 58.0 | 27% |
| 3 | 80.4 | 62.2 | 23% |
| 4 | 81.6 | 66.4 | 19% |
| 5 | 82.8 | 70.6 | 15% |
| 6 | 84.0 | 74.8 | 11% |
| 7 | 85.2 | 79.0 | 7% |
| 8 | 86.4 | 83.2 | 4% |

## 2.1. *Major components*

New feed-through models have been tested on two prototypes with 234 wires. The test results show that the feed-throughs have a good insulated performance, and a long term test for crimping shows no wire broken or tension lost happening for more than three months. Total 12000 feed-throughs have been made and delivered for test.

New endplates are under construction in the manufactory. There are nine steps in each aluminum endplate, and about 2000 φ3.2mm holes with the tolerance of less than 25μm in radius will be drilled on the plate. The assembly tolerance of two endplate is less than 50μm.

The new inner carbon fiber cylinder is also being manufactured. The thickness of the inner cylinder is 1mm to reduce the material budget. The tolerance in diameter and length is less than 80μm, and the deflection in Z direction under stress of 500kg is less than 50μm. The aluminum foils will be pasted on both surfaces of the cylinder for shielding.

### 2.2. *Wire stringing test*

A three-stepped endplate prototype with six layers was built for wire stringing test, due to the increased difficulty caused by the new endplate structure. The prototype has three steps just same as the new inner chamber structure in geometry. Total 120 stereo wires were arranged with the similar cell size. Uniform distribution of wire tension test result proved that wire stringing on the new structure was not a critical problem.

### 3. Simulation

Monte Carlo simulation has been done with the samples of $\pi^+$ track. The simulation results are shown in fig.2 and fig.3. Tracking efficiency changes with momentum and theta, momentum resolution look quite similar for current MDC and MDC with new inner chamber. Because the positions of the feed-throughs are not changed in φ direction, shortening the wire length leads to a larger stereo angle than the current one. As a result, the spatial resolution in Z direction for the new chamber improves a little bit.

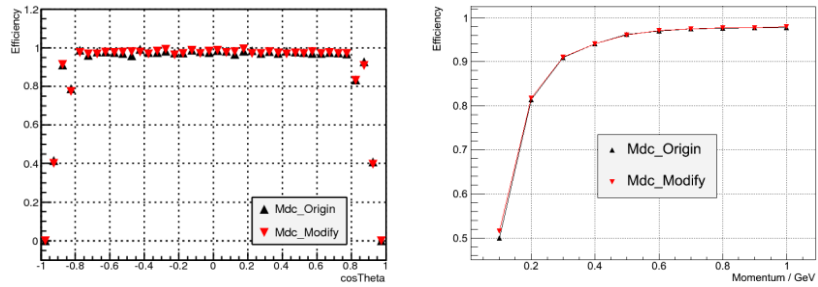

Fig. 2. Tracking efficiency vs. cosθ (left) and momentum (right) compared between current MDC and MDC with new inner chamber.

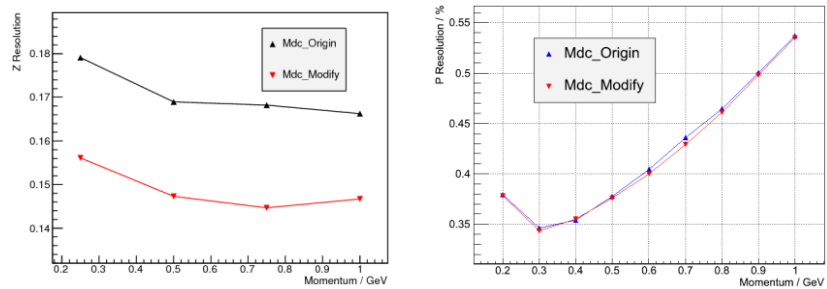

Fig. 3. Spatial resolution in Z direction (left) and momentum resolution (right) compared between current MDC and MDC with new inner chamber.

### 4. Study on inner chamber removal

Because MDC has no outer cylinder for the inner chamber and no inner cylinder for the outer chamber, the inner chamber and the outer chamber are joined together at the endplates. The replacement of the inner chamber is not easy. A test prototype was constructed to simulate the whole replacement procedure, which include the removal of the sealing glue at the endplates, pulling out the inner chamber, and monitoring the wire tension during the operation. Fig.4 shows the experiments of inner chamber removal. From the experiments, we know that the glue can be removed without problems, but very carefully operation is needed to avoid damaging the feed-throughs and the wires. The inner chamber can be removed out with the suitable designed toolings, but stress on inner chamber due to some deformation or external force need to be well considered, so a real-time wire tension measurement is very necessary. Wire tension probably changes a little during the operating but will come back to the normal value after then.

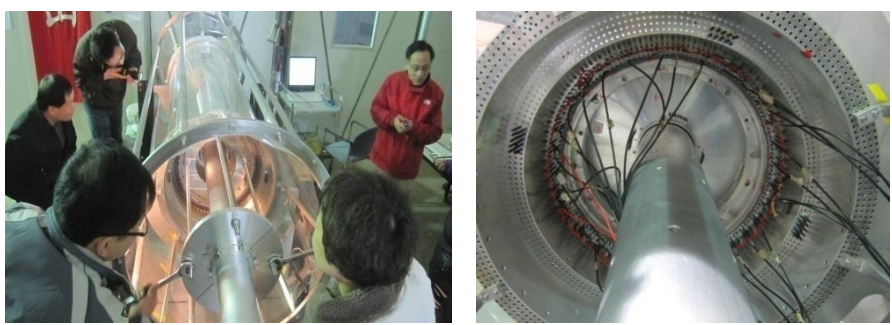

Fig. 4.  The inner chamber removal prototype (left) and wire tension test during the removal operation (right).

## 5. Conclusion

A new inner drift chamber with multi-stepped endplates to shorten the wire length out of effective detection solid angle is designed and under construction, which will reduce about 31% to 4% of the background counting rate for each layer. Although this new structure can not relieve aging effect of each cell, it can reduce the wire dark current, which will be of benefit not only to the resolution in Z direction but also to chamber safe working. The manufactures of the major components are going smoothly. The construction of the new inner drift chamber will be accomplished by the end of April 2014.

A test prototype was built for the inner chamber removal study. We have some experiences from the removal experiments. The inner chamber can be removed out with the suitable toolings safely with very careful operation.

## Acknowledgments

We would like to thank Chinese Academy of Sciences for financial support.

## References

1. BESIII Collaboration, Preliminary design report of the BESIII detector, 2004.
2. C. Chen et al., The BESIII drift chamber, *IEEE Nuclear Science, Symposium Conference Record*, 1844 (2007).